\documentstyle[12pt,graphicx,psfrag]{article}
\input epsf.tex

\newcommand{\beq}{\begin{equation}}
\newcommand{\eeq}{\end{equation}}
\newcommand{\beqa}{\begin{eqnarray}}
\newcommand{\eeqa}{\end{eqnarray}}
\newcommand{\ba}{\begin{array}}
\newcommand{\ea}{\end{array}}
\newcommand{\CR}{\nonumber \\}

\newcommand{\La}{\Lambda}

\begin{document}
\begin{titlepage}
\begin{flushright}
hep-th/yymmnn \\
UTHEP-395 \\
December, 1998
\end{flushright}
\vspace{1.5cm}
\begin{center}
{\Large \bf 
String Junctions and the BPS Spectrum of\\
N=2 SU(2) Theory with Massive Matters}
\lineskip .75em
\vskip2.5cm
{\large Yukiko Ohtake}
\vskip 1.5em
{\large\it Institute of Physics, University of Tsukuba \\
Ibaraki 305-8571, Japan}
\end{center}
\vskip3cm
\begin{abstract}
We study the BPS spectrum of four dimensional $N=2$ $SU(2)$ theory with 
massive fundamental matters using the D3-brane probe.
Since the BPS states are realized by string webs 
subject to the BPS conditions, 
we determine explicitly the configurations of such webs.
It is observed that there appear BPS string webs with multiple of junctions
corresponding to the fact that the curves of marginal 
stability in massive theory are infinitely nested.
In terms of the string configurations, 
various properties of the curves of marginal stability are explained 
intuitively.
\end{abstract}
\end{titlepage}
\baselineskip=0.7cm

\newpage

\section{Introduction}

It is well known that the non-perturbative BPS spectrum of four-dimensional
$N=2$ supersymmetric gauge theory is determined by the celebrated 
Seiberg-Witten formula \cite{SeWi}.
As the simplest case the BPS spectrum in $SU(2)$ theory with $N_f$ fundamental
matters has been studied in great detail \cite{BF2}\cite{BF3}.
On the other hand, the recent analysis based on the D3-brane probe approach 
has provided the stringy realization of the $SU(2)$ BPS states
\cite{Sen}-\cite{MNS}.
For instance, quark hypermultiplets in $N_f=4$ theory are realized as open 
strings connecting a probe D3-brane with a background D7-brane,
while the W-boson as a string junction.
An amusing aspect of this approach is that the subtle properties of the 
curves of marginal stability of BPS states in the moduli space are understood 
rather intuitively.

The D3-brane probe analysis of the $SU(2)$ BPS states has been performed so far 
in pure gauge theory ($N_f=0$) and gauge theory with $N_f=2,3$ massless 
matters.
In these examples the moduli space contains two singularities. 
Our purpose in this paper is first to extend the analysis to the case of 
$N_f=1$ massless matter for which there appear three singularities in the 
moduli space.
This will be done in sect.2 where we also review the D3-probe realization 
of $N=2$ $SU(2)$ theory with fundamental matters.
We then proceed to the theory with $N_f=2$ massive matters in sect.3.
Finally sect.4 is devoted to our conclusions.

\section{$SU(2)$ theory with $N_f=1$ massless matter}

We begin with $SU(2)$ theory with $N_f=1$ massless matter.
Denoting by $z$ the order parameter of the Coulomb branch, 
it was shown that the theory has three singularities at $z=z_i$ $(i=1,2,3)$ 
in the moduli space
where a dyon (or monopole) with electric-magnetic charges $(p_i,q_i)$ becomes
massless. 
In the type IIB setup the same theory is realized as the world volume
theory of a D3-brane in the background with parallel $(p_i,q_i)$ 7-branes
at $z=z_i$.
Here $z$ is the coordinate of the complex plane $z=x^8+ix^9$,
$p_i$ and $q_i$ label the R-R and NS-NS charges of the 7-brane.
Following the conventions in \cite{BF2},
we fix the background with a $(2,-1)$ 7-brane at $z_1=-1$,
a $(-1,1)$ 7-brane at $z_2=e^{i\frac{\pi}{3}}$ and
a $(0,1)$ 7-brane at $z_3=e^{-i\frac{\pi}{3}}$.
The singularities cause the monodromy on the moduli space which is 
represented by the $SL(2,{\bf Z})$ matrices,
\beq
M_i=\left(\ba{cc} 1-p_iq_i & p_i^2\\-q_i^2&1+p_iq_i\ea\right).
\label{eqn:monodromy}
\eeq
In the IIB description it is due to the fact that 7-branes are the magnetic 
sources for $\tau$ where $\tau=\chi+ie^{-\phi}$ with $\chi$ being a R-R 
scalar and $\phi$ a dilaton.
Thus going around the $(p_i,q_i)$ 7-brane anti-clockwise
$\tau$ undergoes a monodromy (\ref{eqn:monodromy}).
We introduce the branch cuts emanating from the 7-branes on the $z$-plane 
specified as in \cite{BF2}. 
The moduli parameter (usually denoted as $u$) represents the position of the
D3-brane.

A $(p,q)$ dyonic state of $SU(2)$ theory is realized by a $(p,q)$ string 
ending on the D3-brane, 
where $p$ and $q$ now label the NS-NS and R-R charges of the string.
To have a finite mass, the string must end on the $(p_i,q_i)$ 7-branes.
This is allowed when the string separates to $n_i$ $(p_i,q_i)$ strings 
with $(p,q)=\sum_in_i(p_i,q_i)$ making a string web,
or to strings related to $(p_i,q_i)$ strings by crossing the branch cuts.
The latter configuration can be deformed continuously to the former
by virtue of the Hanany-Witten effect \cite{GZ},
and hence we characterize the family of such string webs by $n_i$.

In order for the string webs to be BPS, the possible values of $n_i$
are restricted.
In the framework of M-theory a string web is lifted to a membrane which
must be a smooth holomorphic curve $J$ to conserve a half of 
the supersymmetries \cite{KL}-\cite{KS}.
Then its self-intersection number is given by
\beq
(J\cdot J)=2g-2+b,
\eeq
where $g$ is the genus and $b$ is the number of boundaries of $J$ \cite{dWHIZ}.
We consider $J$ with $b\geq 1$ because they must have boundaries on an M5-brane
to which the D3-brane probe is lifted under T-duality.
On the IIB side the intersection number is expressed in terms of $n_i$
\cite{dWZ}, 
\beq
(J\cdot J)=-\sum_i n_i^2
        +\sum_{i<j}n_in_j\left|\ba{cc}p_i&p_j\\q_i&q_j\ea\right|.
\label{eqn:in}
\eeq
For the background under consideration we thus have
\beq
(J\cdot J)=-\frac{1}{2}
        \left[(n_1+n_2-n_3)^2+(n_1-n_3)^2+n_2^2\right].
\label{eqn:in1}
\eeq
Since $(J\cdot J)\geq -1$ we obtain from (\ref{eqn:in1}) the possible
values of $n_i$ for the BPS string webs. 
The result is presented in Table \ref{tbl:nf=1}.
\begin{table}
\hspace{4cm}
\begin{tabular}{|c|c|c|c|}\hline
        &$b$    &$(n_1,n_2,n_3)$        &$(p,q)$\\ \hline
(i)     &$2$    &$\pm(1,0,1)$           &$\pm(2,0)$\\
(ii)    &$1$    &$\pm(n,1,n)$           &$\pm(2n+1,-1)$\\
(iii)   &$1$    &$\pm(n,0,n+1)$         &$\pm(2n,1)$\\
(iv)    &$1$    &$\pm(n,1,n+1)$         &$\pm(2n+1,0)$\\ \hline
\end{tabular}
\caption{The possible values of $n_i$ for BPS string webs in $SU(2)$ theory 
with $N_f=1$ massless matter.}
\label{tbl:nf=1}
\end{table}

Note that given a value of $(n_1,n_2,n_3)$ the configuration of 
corresponding string web is not uniquely determined.
Thus our next step is to determine the BPS string web explicitly on 
the $z$-plane by requiring their masses to be minimum.
The mass of a $(p,q)$ string stretched along 
the curve $C$ originating from $z=z_0$ is given by the Seiberg-Witten 
solutions $a(z)$ and $a_{D}(z)$ \cite{Sen3},
\beq
m=\int_{C} \left| pda(z)-qda_{D}(z) \right|
 \geq \left| \int_{C} \left( pda(z)-qda_{D}(z) \right)\right|.
\label{eqn:mass}
\eeq
Then, for the BPS string, the curve $C$ is taken to be
\beq
\mbox{arg}\left[pa(z)-qa_{D}(z)-pa(z_0)+qa_{D}(z_0)\right]
=\phi,
\label{eqn:geodesic}
\eeq
where $\phi$ is a constant between 0 and $2\pi$ \cite{BFy}\cite{MNS}.

Suppose that a $(p,q)$ sting emanating from $z=z_0$ and a $(p',q')$ string 
from $z=z_0'$ meet at a point $z=\tilde{z}$ which is a vertex of 
certain string web.
The curves associated with these BPS strings have the common argument $\phi$
since the mass of a web is simply given by the sum of the masses of component
strings.
Note that when we construct a web with the argument $\phi$
we can use BPS strings with $\phi+\pi$ as well
because (\ref{eqn:geodesic}) for a $(p,q)$ string with $\phi+\pi$
is considered as the equation for a $(-p,-q)$ string with $\phi$,
i.e. a $(p,q)$ string whose direction is opposite to the string with
$\phi$.
In either case we find
\beq
\mbox{Im}\frac{pa(\tilde{z})-qa_{D}(\tilde{z})-pa(z_0)+qa_{D}(z_0)}
{p'a(\tilde{z})-q'a_{D}(\tilde{z})-p'a(z_0')+q'a_{D}(z_0')}=0.
\label{eqn:BPS}
\eeq
We note that this is nothing but the condition which determines the curve of 
marginal stability (CMS) in $SU(2)$ theory where the BPS spectrum changes.

\begin{figure}
\hspace{1.7cm}
\psfrag{Z1}[][]{$z_1$}
\psfrag{Z2}[][]{$z_2$}
\psfrag{Z3}[][]{$z_3$}
\psfrag{Z1st}[][]{$(1,0)$}
\psfrag{Z2st}[][]{$(1,-1)$}
\psfrag{Z3st}[][]{$(0,1)$}
\psfrag{Z4st}[][]{$(0,-1)$}
\psfrag{Z5st}[][]{$(-1,1)$}
\psfrag{Z6st}[][]{$(-2,1)$}
\psfrag{TT1}[][]{(A)}
\psfrag{TT2}[][]{(B)}
\includegraphics[width=12cm]{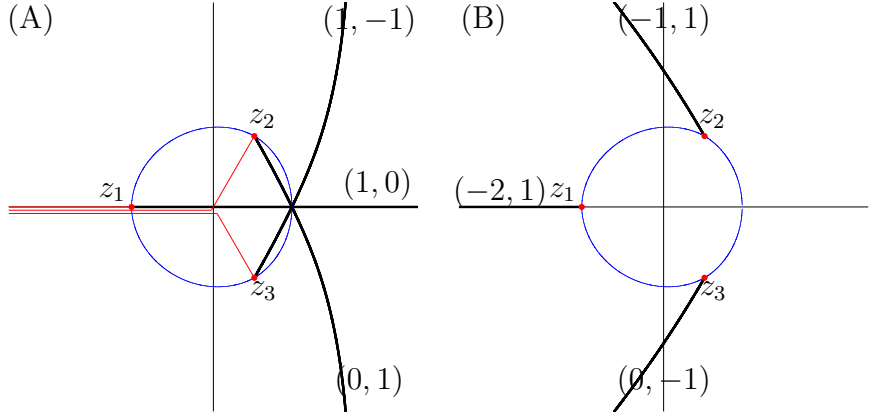}
\caption{The BPS strings from the 7-branes with $\phi=0$ and CMS for $N_f=1$.
Dotted lines toward $z\rightarrow -\infty$ are the branch cuts.}
\label{fig:nf=1}
\end{figure}

Setting $\phi=0$ and using the explicit forms of $a(z)$ and $a_D(z)$ in 
(\ref{eqn:aad1}), we have the BPS strings emanating from 
the 7-branes. These strings are depicted in Fig.\ref{fig:nf=1}A.
In addition there are the strings with opposite directions as shown in 
Fig.\ref{fig:nf=1}B.
They do not meet other strings and cannot construct junctions.
On the other hand we observe that the strings in Fig.\ref{fig:nf=1}A meet 
at a point.
The point turns out to be a vertex {\bf v} of a 3- or 4-string 
junction when another BPS string with charge $(p,q)$ listed 
in Table \ref{tbl:nf=1} is attached at {\bf v}.
We see from (\ref{eqn:BPS}) that {\bf v} is on a CMS with
$\mbox{Im}a_D(z)/a(z)=0$ for any $\phi$ since $p_ia(z_i)-q_ia_D(z_i)=0$.
The $(p,q)$ string goes outside of the CMS because of the balance of tensions 
\cite{Sen2}.
The configuration is deformed continuously when we vary $\phi$.
As $\phi$ increases, {\bf v} moves anti-clockwise on the CMS 
and the $(p,q)$ string sweeps outside the CMS.
The Hanany-Witten effect occurs at $\phi=(2j-1)\pi/6$ when {\bf v}
is located at $z=z_j$. 

In considering string junctions, we note that the strings listed in 
Table \ref{tbl:nf=1}(iv) are not all independent 
because the configurations of them and $\pm(2,0)$ string obey the identical 
equation as seen from (\ref{eqn:geodesic}).
We can then choose $\pm(1,0)$ string from the 7-brane at $z=z_j$ as an 
independent string for $\frac{1-2j}{3}\pi<\mbox{arg}z<\frac{3-2j}{3}\pi$, 
while the others are considered as the $\pm(1,0)$ string connected with
$\pm(2,0)$ webs on the boundary.

If a $(p,q)$ string can end on the D3-brane introduced as a probe
there exists a $(p,q)$ BPS state in $SU(2)$ theory.
We see that there exist the $(p,q)$ BPS states listed in Table \ref{tbl:nf=1}
(i)-(iii) and $(1,0)$ in (iv) outside the CMS.
Inside the CMS, the allowed BPS state is represented by a string connecting 
the D3-brane with a 7-brane.
These reproduce the known BPS spectrum obtained from the field theoretical 
analysis.

\section{$SU(2)$ theory with $N_f=2$ massive matters}

Next we consider $SU(2)$ theory with $N_f=2$ massive fundamental matters. 
We assume for simplicity that the matter multiplets have the common mass
$m\in{\bf R}$.
The mass formula for a $(p,q)_s$ BPS state is given by \cite{SeWi}
\[
m_{\mbox{{\tiny BPS}}}=\left|pa(u)-qa_D(u)-\frac{ms}{\sqrt{2}}\right|,
\]
where the quantum number $s$ is associated with the conserved fermion number.
In comparing with (\ref{eqn:mass})(\ref{eqn:geodesic})
we have $pa(z_0)-qa_D(z_0)=ms/\sqrt{2}$.
There appear three singularities in the moduli space.
The structure of the moduli space changes at $m=\La/2$ where the 
two singularities collide each other. 
We shall consider the cases $m<\La/2$ and $m>\La/2$ separately.

In the case $m<\La/2$,
following the conventions in \cite{BF3},
we consider the background with a $(1,-1)_{-1}$ 7-brane at 
$z_1=-\frac{1}{8}\La(\La+8m)$, 
a $(1,-1)_1$ 7-brane at $z_2=-\frac{1}{8}\La(\La-8m)$ and 
two $(0,1)_0$ 7-branes at $z_3=z_4=\frac{1}{8}(\La^2+8m^2)$. 
The branch cuts stretch from $z=z_i$ to $-\infty$ along the real axis. 
The self-intersection number is obtained from (\ref{eqn:in})
\[
(J\cdot J)=-\frac{1}{2}\left[
        (n_1-n_3)^2+(n_1-n_4)^2+(n_2-n_3)^2+(n_2-n_4)^2\right],
\]
from which we read off the possible values of $n_i$ for the BPS string webs,
see Table \ref{tbl:nf=2s}.
A pair of string webs characterized by $(n_1,n_2,n_4,n_3)$ and
$(n_1,n_2,n_3,n_4)$ forms a doublet of the unbroken $SU(2)$ flavor symmetry. 
They have the same BPS configurations.
\begin{table}
\hspace{2.3cm}
\begin{tabular}{|c|c|c|c|}\hline
        &$b$    &$(n_1,n_2,n_3,n_4)$ or $(n_1,n_2,n_4,n_3)$
                                        &$(p,q)$\\\hline
(i)     &$2$    &$\pm(1,1,1,1)$         &$\pm(2,0)_0$\\
(ii)    &$1$    &$\pm(n,n,n,1)$         &$\pm(2n,1)_0$\\
(iii)   &$1$    &$\pm(n+1,n,n,n)$       &$\pm(2n+1,-1)_{-1}$\\
(iv)    &$1$    &$\pm(n,n+1,n,n)$       &$\pm(2n+1,-1)_1$\\
(v)     &$1$    &$\pm(n,n+1,n,n+1)$     &$\pm(2n+1,0)_1$\\\hline
\end{tabular}
\caption{The possible values of $n_i$ for BPS string webs in 
$SU(2)$ theory with massive $N_f=2$ ($m<\frac{\La}{2}$).}
\label{tbl:nf=2s}
\end{table}

\begin{figure}
\psfrag{Z1}[][]{$z_1$}
\psfrag{Z2}[][]{$z_2$}
\psfrag{Z3}[][]{$z_3$}
\psfrag{Z1st}[][]{$(1,-1)_{-1}$}
\psfrag{Z2st}[][]{$(1,-1)_1$}
\psfrag{Z3st}[][]{$(0,1)_0$}
\psfrag{TT1}[][]{(A) $\phi=\pi/4$}
\psfrag{TT2}[][]{(B) $\phi=7\pi/16$}
\psfrag{TT3}[][]{(C) $\phi=9\pi/16$}
\psfrag{TT4}[][]{(D) $\phi=3\pi/4$}
\includegraphics[width=15.5cm]{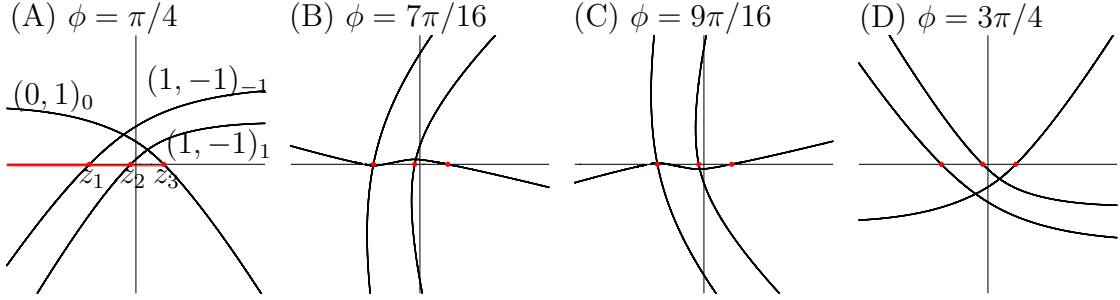}
\caption{BPS strings for $\alpha_i<\phi<\alpha_{i+1}$ $(i=0,1,2,3)$
with $\alpha_0=0$ and $\alpha_4=\pi$.
We set $m/\La=0.1$ for which $\alpha_1=0.43\pi$, $\alpha_2=0.5\pi$
and $\alpha_3=0.57\pi$.}
\label{fig:HWs}
\end{figure}

A BPS string web containing $k$ vertices is constructed from  
a $k_1$-vertex web and a $k_2$-vertex web where $k_1+k_2= k-1$.
When BPS strings with some $\phi$ are given,
we can construct BPS string webs successively from small $k$. 
First, numerical calculations enable us to determine the explicit forms 
of 0-vertex webs, i.e. the BPS strings ending on the 7-branes.
The strings may hit a mutually non-local 7-brane at $\phi=\alpha$ 
where $\alpha$ depends on the value of $m/\La$.
Then the configuration changes depending upon $\phi<\alpha$ or not.
Moreover numerical calculations give us four types of configurations 
depending on $\alpha$ as shown in Fig.\ref{fig:HWs} where we set $m/\La=0.1$
and have used the explicit forms of $a(z)$ and $a_D(z)$ in (\ref{eqn:aad2}).
We present the strings stretched in the opposite directions
in the same picture which may contribute when string webs are constructed.
For $\phi>\pi$, the configurations are the same with those of $\phi-\pi$
with charges $(-p,-q)_{-s}$.

\begin{figure}
\psfrag{N0}[][]{0}
\psfrag{N1}[][]{1}
\psfrag{N2}[][]{2}
\psfrag{N3}[][]{3}
\psfrag{N4}[][]{4}
\psfrag{TTa}[][]{(a)}
\psfrag{TTb}[][]{(b)}
\psfrag{TTc}[][]{(c)}
\psfrag{(pa,qa)}[][]{$(1,-1)_{-1}$}
\psfrag{(pb,qb)}[][]{$(1,-1)_1$}
\psfrag{(pc,qc)}[][]{$(0,1)_0$}
\psfrag{TTd}[][]{(d)}
\psfrag{TTe}[][]{(e)}
\psfrag{C0p}[][]{$C_{0}^{+}$}
\psfrag{l1=d}[][]{$l_1=1$}
\psfrag{l2=d}[][]{$l_2=2$}
\psfrag{l1=e}[][]{\hspace{-1mm}$l_1=1$}
\psfrag{l2=e}[][]{\hspace{-1mm}$l_2=1$}
\psfrag{(pd,qd)}[][]{$(1,1)_{-1}$}
\psfrag{(pe,qe)}[][]{$(1,0)_{-1}$}
\psfrag{TTf}[][]{(f)}
\psfrag{TTg}[][]{(g)}
\psfrag{C0m}[][]{$C_{0}^{-}$}
\psfrag{l1=f}[][]{\hspace{0.5mm}$l_1=1$}
\psfrag{l2=f}[][]{\hspace{0.5mm}$l_2=2$}
\psfrag{l1=g}[][]{\hspace{-1mm}$l_1=1$}
\psfrag{l2=g}[][]{\hspace{-1mm}$l_2=1$}
\psfrag{(pf,qf)}[][]{$(1,1)_{1}$}
\psfrag{(pg,qg)}[][]{$(1,0)_{1}$}
\psfrag{TTh}[][]{(h)}
\psfrag{TTj}[][]{(j)}
\psfrag{TTk}[][]{(k)}
\psfrag{Ci}[][]{$C^{\infty}$}
\psfrag{l1=h}[][]{\hspace{2mm}$l_1=n$}
\psfrag{l2=h}[][]{\hspace{9mm}$l_2=n+1$}
\psfrag{l1=j}[][]{$l_1=1$}
\psfrag{l2=j}[][]{$l_2=1$}
\psfrag{l1=k}[][]{\hspace{7mm}$l_1=n+1$}
\psfrag{l2=k}[][]{\hspace{-0.5mm}$l_2=n$}
\psfrag{(ph,qh)}[][]{$(2n+1,1)_{1}$}
\psfrag{(pj,qj)}[][]{$(2,0)_0$}
\psfrag{(pk,qk)}[][]{$(2n+1,-1)_{-1}$\hspace{8mm}}
\psfrag{TTl}[][]{(l)}
\psfrag{TTm}[][]{(m)}
\psfrag{Cm2m}[][]{$C_{-2}^{-}$}
\psfrag{l1=l}[][]{\hspace{1mm}$l_1=1$}
\psfrag{l2=l}[][]{\hspace{1mm}$l_2=2$}
\psfrag{l1=m}[][]{$l_1=1$\hspace{2mm}}
\psfrag{l2=m}[][]{$l_2=1$\hspace{2mm}}
\psfrag{(pl,ql)}[][]{$(3,-1)_{1}$}
\psfrag{(pm,qm)}[][]{$(2,-1)_{0}$}
\psfrag{TTn}[][]{(n)}
\psfrag{TTo}[][]{(o)}
\psfrag{C2p}[][]{$C_{2}^{+}$}
\psfrag{l1=n}[][]{$l_1=1$}
\psfrag{l2=n}[][]{$l_2=1$}
\psfrag{l1=o}[][]{$l_1=2$\hspace{2mm}}
\psfrag{l2=o}[][]{$l_2=1$\hspace{2mm}}
\psfrag{(pn,qn)}[][]{$(2,1)_{0}$}
\psfrag{(po,qo)}[][]{\hspace{2mm}$(3,1)_{-1}$}
\psfrag{TTp}[][]{(p)}
\psfrag{TTq}[][]{(q)}
\psfrag{C2np}[][]{$C_{2n+2}^{+}$}
\psfrag{l1=p}[][]{$l_1=1$}
\psfrag{l2=p}[][]{$l_2=1$}
\psfrag{l1=q}[][]{$l_1=2$\hspace{2mm}}
\psfrag{l2=q}[][]{$l_2=1$\hspace{2mm}}
\psfrag{(pp,qp)}[][]{$(2n+2,1)_{0}$\hspace{8mm}}
\psfrag{(pq,qq)}[][]{$(2n+3,1)_{-1}$\hspace{8mm}}
\psfrag{TTr}[][]{(r)}
\psfrag{TTs}[][]{(s)}
\psfrag{Cm2np}[][]{$C_{-2n-2}^{-}$}
\psfrag{l1=r}[][]{$l_1=1$}
\psfrag{l2=r}[][]{$l_2=1$}
\psfrag{l1=s}[][]{$l_1=2$\hspace{1mm}}
\psfrag{l2=s}[][]{$l_2=1$\hspace{1mm}}
\psfrag{(pr,qr)}[][]{\hspace{8mm}$(2n+3,-1)_{1}$}
\psfrag{(ps,qs)}[][]{\hspace{4mm}$(2n+2,-1)_{0}$}
\psfrag{l1(a)+}[][]{$l_1$(a)$+$}
\psfrag{l2(c)}[][]{$l_2$(c)}
\psfrag{l1(b)+}[][]{$l_1$(b)$+$}
\psfrag{l2(f)}[][]{$l_2$(f)}
\psfrag{l2(g)}[][]{$l_2$(g)}
\psfrag{l1(e)+}[][]{$l_1$(e)$+$}
\psfrag{l2(h)}[][]{$l_2$(h)}
\psfrag{l1(g)+}[][]{$l_1$(g)$+$}
\psfrag{l2(k)}[][]{$l_2$(k)}
\includegraphics[width=16cm]{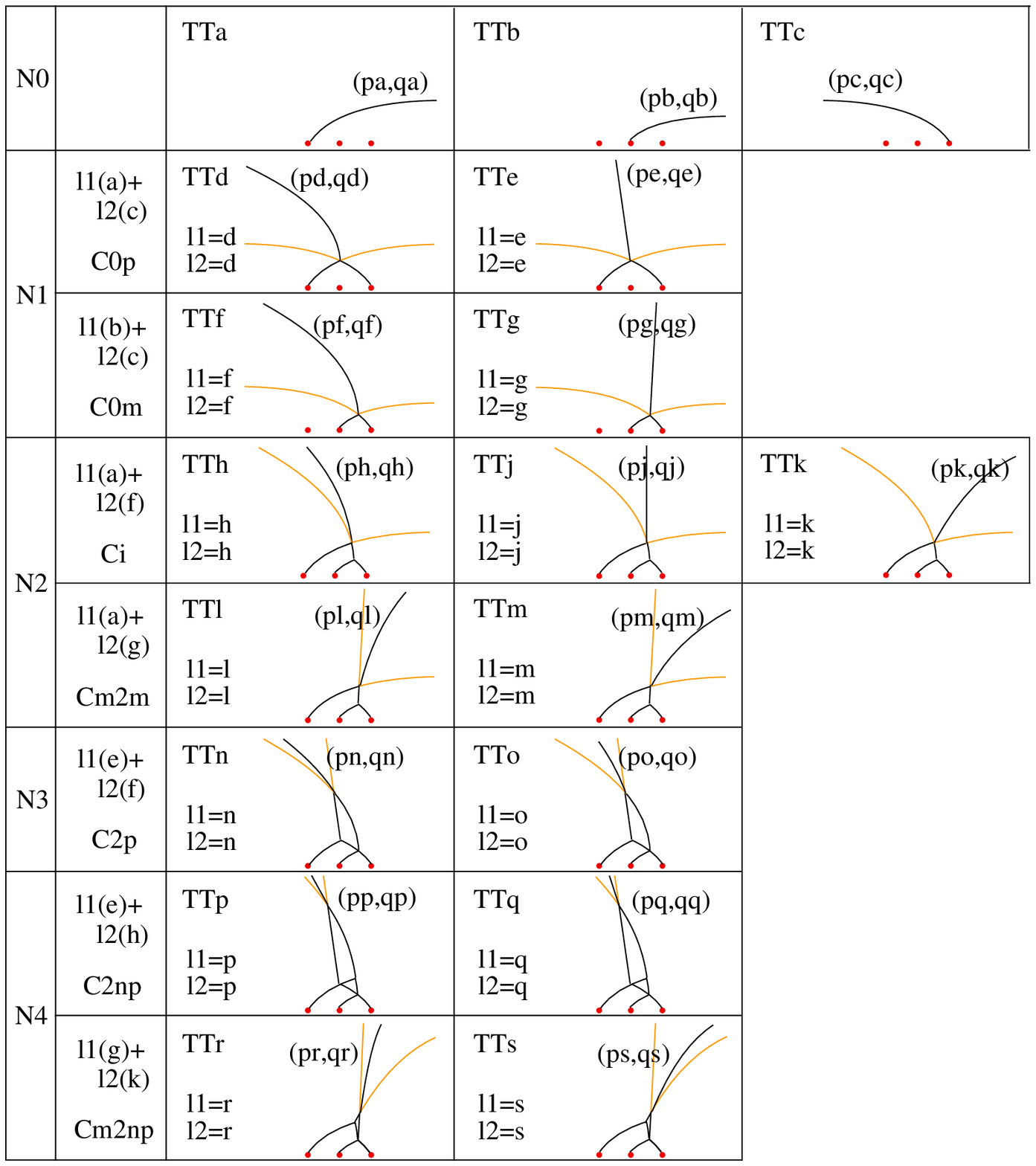}
\vspace{-8mm}
\caption{The BPS string webs of $N_f=2$ where $0<\phi<0.43\pi$ and 
$m/\La=0.1$. For the figures h,k,p-s we set $n\geq 1$. 
The number of vertices is shown in the first column.
In the second column we show the component webs and CMS.
For instance the web (d) is made out of $l_1=1$ string of (a) and $l_2=2$
strings of (c).
The direction of the outgoing string is restricted by the directions of
component webs depicted by dotted lines.
It is also bounded from the left (or right) by the outgoing string of the web
whose configuration is presented in the left (or right) figure.}
\label{fig:sm}
\end{figure}

From this observation, one may inspect configurations of webs with higher
vertices without doing explicit computations.
For example, let us consider the case as in Fig.\ref{fig:HWs}A.
The strings possible to make vertices are given in Fig.\ref{fig:sm}a-c. 
The strings with the same charge $(p,q)$ cannot cross 
since they are parallel under the metric with respect to which 
(\ref{eqn:mass}) is the geodesic condition. 
Then the strings in Fig.\ref{fig:sm}a and Fig.\ref{fig:sm}b cannot 
construct a string junction 
while the string webs with a vertex are constructed by $l_1$ strings in
Fig.\ref{fig:sm}a and $l_2$ strings in Fig.\ref{fig:sm}c
or $l_1$ strings in Fig.\ref{fig:sm}b and $l_2$ strings in Fig.\ref{fig:sm}c.
We see that $(l_1,l_2)=(1,2)$ and $(1,1)$ respectively, since
the possible values of $l_i$ are determined from Table \ref{tbl:nf=2s}
and $(n_1,n_2,n_3+n_4)=(l_1,0,l_2)$ or $(0,l_1,l_2)$.
Note that relative positions of outgoing prongs are determined by comparing 
the values of $|l_1/l_2|$.
As it decreases the direction of the outgoing prong become more parallel
to the direction of the strings in Fig.\ref{fig:sm}c.
This consideration becomes more relevant than the balance of tension
to restricting the directions of outgoing strings.
Using these restrictions 
we are now able to successively construct $k$-vertex string webs with $k=2,3$
and 4 (see Fig.\ref{fig:sm}h-s).
Note that they consist of 3-string junctions.
The configurations which look like 4-string junctions in 
Fig.\ref{fig:sm}n-s separate to two types of 3-string junctions.
Examining further we see that $k$-vertex string webs with $k\geq 5$ cannot be
constructed.

We have drawn Fig.\ref{fig:sm} corresponding to Fig.\ref{fig:HWs}A.
(As long as $\alpha_0<\phi<\alpha_1$ the figures are essentially the same.)
We obtain the configurations of string webs corresponding to
Fig.\ref{fig:HWs}B-D in the same way.
It is seen that the BPS string webs of type (v) in Table \ref{tbl:nf=2s} 
with $n\neq 0,-1$ do not appear in any $\phi$ in agreement with the field 
theory.
For the string webs whose outermost vertices are in the 
upper half-plane,
the results are understood from Fig.\ref{fig:sm} by considering the ordinary 
Hanany-Witten effect. 
In fact, for $0.43\pi<\phi<0.5\pi$ in Fig.\ref{fig:HWs}B,
the string junction $(1,0)_{-1}$ in Fig.\ref{fig:sm}e is replaced by a 
string from $z=z_3$,
according to which the corresponding junctions in Fig.\ref{fig:sm}n-q are 
also replaced by the strings
while the configurations in Fig.\ref{fig:sm}f-m are essentially unchanged.
These are understood from the Hanany-Witten effect at $\phi=0.43\pi$
due to the 7-brane at $z=z_1$.
For $0.5\pi<\phi<0.57\pi$ in Fig.\ref{fig:HWs}C, the string junction  
$(1,0)_1$ in Fig.\ref{fig:sm}g is replaced by a string from $z=z_3$ 
according to which the corresponding junctions in Fig.\ref{fig:sm}l-m are also 
replaced and the other string webs cannot be constructed.
These are understood from the Hanany-Witten effect at $\phi=0.5\pi$
due to the 7-brane at $z=z_2$.

Taking into account the string webs whose vertices are in the 
lower plane, we find that the Hanany-Witten effect occurs in a more 
complicated fashion.
For instance, comparing the result from Fig.\ref{fig:HWs}A and 
Fig.\ref{fig:HWs}D the configuration of $(p,q)_s$ string web looks like
the mirror image of the configuration of $(-p,q)_{-s}$ string web
with respect to the real axis in the $z$-plane.
Then in the limit $\phi=0$, the vertices of the strings move to the lower 
half-plane and the configurations change significantly.
As an example the string web in Fig.\ref{fig:sm}m changes to the
configuration which is the mirror image of Fig.\ref{fig:sm}n.
We note that they are exact mirror images.
To understand this, note that the Seiberg-Witten curve of the theory with 
$m\in{\bf R}$, i.e. $y^2=x^3+f(z)x+g(z)$ from which $a(z)$ and $a_D(z)$ are 
derived, contains no complex parameters except for $z$.
Then $\left[a(z)\right]^{*}$ and $\left[a_D(z)\right]^{*}$ derived 
from the curve which is the complex conjugate of the Seiberg-Witten curve 
$y^2=x^3+f(\bar{z})x+g(\bar{z})$ are equal to
$a(\bar{z})$ and $a_D(\bar{z})$ up to the $SL(2,{\bf Z})$ transformation, 
the convention of branch cuts and phases.
This leads $\left[a(z)\right]^{*}=a(\bar{z})$ and $\left[a_D(z)\right]^{*}
=-a_D(\bar{z})$ except for $z$ on the branch cuts in the conventions we use,
and hence the configuration of $(p,q)_s$ string web with the argument $\phi$ 
is the mirror image of the configuration of $(-p,q)_{-s}$ with the argument 
$\pi-\phi$ .

Let us now discuss the CMS in massive $N_f=2$ theory.
It is known that there exist infinitely many types of CMS.
They are labeled by \cite{BF3},
\beq
C^{\infty}:\mbox{Im}\frac{a(z)}{\epsilon a_D(z)+\frac{m}
                {\sqrt{2}}}=0,\hspace{3mm}
C^{\pm}_n:\mbox{Im}\frac{a(z)\pm\frac{m}{\sqrt{2}}}{\epsilon a_D(z)-na(z)}=0,
\label{eqn:CMS}
\eeq
where $\epsilon=\mbox{sign}(\mbox{Im}z)$.
The CMS for each BPS state is listed in the second column in Fig.\ref{fig:sm}
($0<\phi<0.43\pi, m/\La=0.1$)
which is determined from the configuration of the corresponding string web
and (\ref{eqn:BPS}).
As we vary $\phi$ the Hanany-Witten effect may occur to change the web
configuration, as a result of which CMS will also change.
This is not the case, however, for the web with the outermost vertex located
in the upper half-plane.
Then, in the lower half-plane, CMS are determined from the mirror relation.
The CMS determined in this way are in agreement with the results of field
theory.

\begin{figure}
\hspace{1cm}
\epsfxsize=13cm
\epsfbox{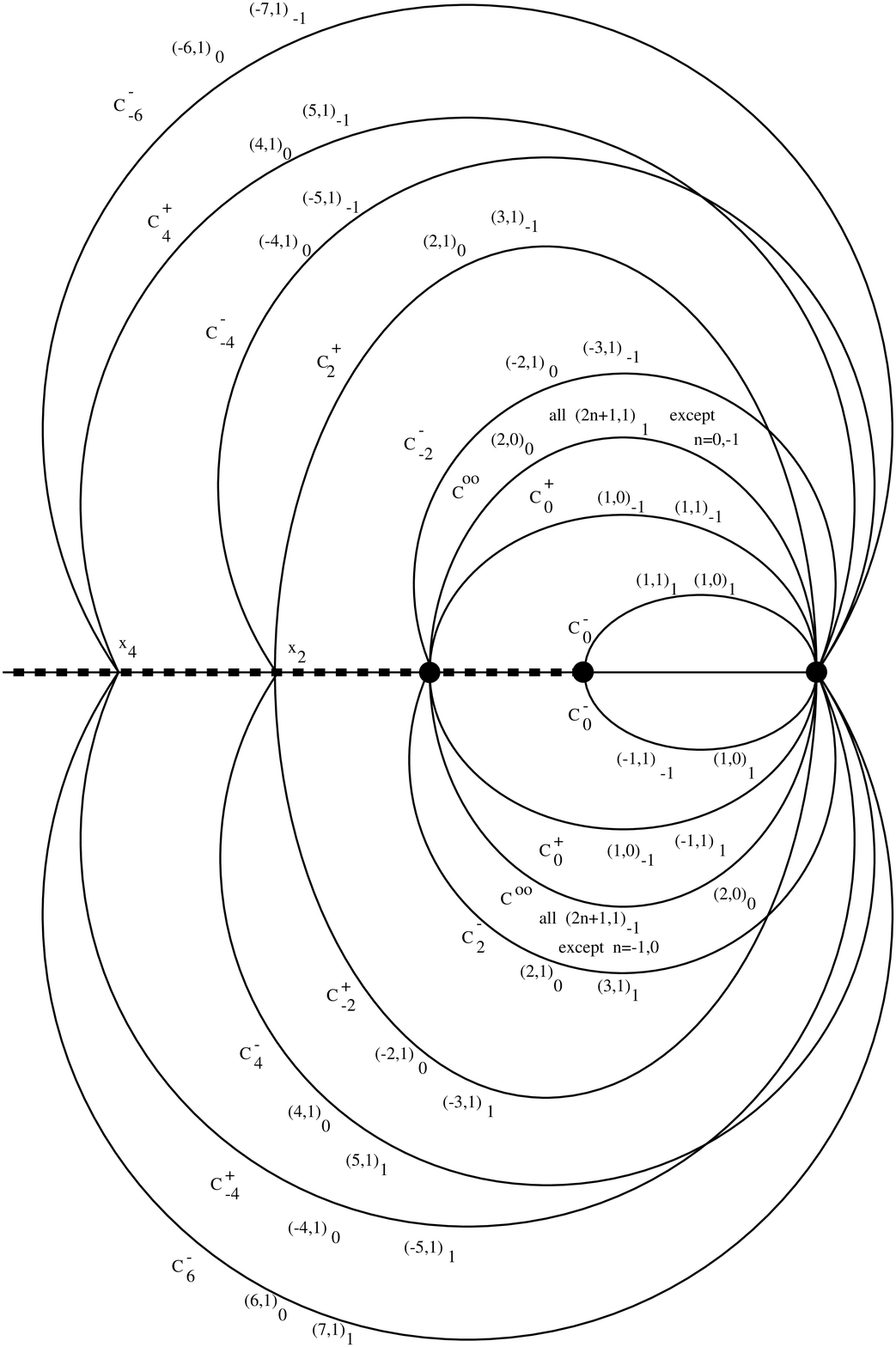}
\caption{CMS for $SU(2)$ theory with $N_f=2$ massive matters depicted 
by Bilal and Ferarri.}
\label{fig:CMS}
\end{figure}

The configurations of CMS for $m<\La/2$ are shown in Fig.\ref{fig:CMS}
as first sketched by Bilal and Ferrari in \cite{BF3}.
One notes the following features of CMS;
(I) BPS states with different charges $(p,q)_s$ can share the same CMS,
(II) there appear variety of the sizes of CMS depending on $(p,q)_s$,
and (III) different CMS merge at some points on the real axis.

First let us look at the string webs in Fig.\ref{fig:sm}h.
It is clearly seen that a prong attached on the vertex on the CMS $C^{\infty}$
can have the charges $(2n+1,1)_1$ with $n\in {\bf Z}$ except for $n=0,-1$.
This is exactly the feature (I) of $C^{\infty}$.
The feature (I) of the other CMS are understood similarly.

Second, the relative size of CMS is determined after solving (\ref{eqn:CMS})
explicitly in the field theory approach.
On the other hand, the result may be geometrically understood when we draw the 
corresponding web configurations in the brane approach.
For example, it is obvious that the CMS $C_0^+$ in Fig.\ref{fig:sm}d-e 
is larger than 
$C_0^-$ in Fig.\ref{fig:sm}f-g.
In this sense we understand the relative size of CMS in a visual way.

Finally we examine the feature (III) of CMS in view of the brane analysis.
We see that all CMS in (\ref{eqn:CMS}) touch the point $z=z_3$ as in 
Fig.\ref{fig:CMS}
since $a_D(z_3)=0$ and $\mbox{Im}a(z)=0$ on the real axis with $z\geq z_3$ due 
to $\left[a(z)\right]^{*}=a(\bar{z})$.
We also see from (\ref{eqn:BPS}) that CMS of the webs depicted in 
Fig.\ref{fig:sm}d-m terminate at the singularities $z=z_1$ or $z_2$.
Since the web of $(p,q)_s$ in the upper $z$-plane is the mirror image 
of the web of $(-p,q)_{-s}$ in the lower $z$-plane,
it is obvious that CMS of $(p,q)_s$ in the upper plane meets CMS 
of $(-p,q)_{-s}$ in the lower half-plane on the real axis.
That CMS $C_{2n}^{+}$ and $C_{-2n-2}^{-}$ meet at the same point on 
the real axis is understood as follows.
The strings $(2,0)_0$ must be the mirror image of $(2,0)_0$, and thus there 
exists, at some $\phi=\alpha$, a $(2,0)_0$ prong stretched the real axis.
Then the $(2n+2,-1)_0$ and $(2n+3,-1)_1$ strings in Fig.\ref{fig:sm}r-s 
whose directions are bounded by the $(2,0)_0$ string from the left 
are placed on the upper half-plane from the vertex at $\phi=\alpha$.
On the other hand, the $(2n+2,1)_0$ and $(2n+3,1)_{-1}$ strings in 
Fig.\ref{fig:sm}n-q bounded from the right are placed on the lower 
half-plane from the vertex.
As $\phi$ moves, suppose that the vertices hit the real axis at $z=x_{2n}$ for 
the $(2n+2,-1)_0$ and $(2n+3,-1)_1$ strings and at $z=x'_{2n}$ for the 
$(2n,1)_0$ and $(2n+1,1)_{-1}$ strings. 
After the Hanany-Witten effect, the $(2n+2,-1)_0$ and $(2n+3,-1)_1$ strings 
change to the strings from the lower half-plane across the cuts.
The charges of the strings before crossing the cuts are $(-2n,1)_0$ and
$(-2n-1,1)_{1}$ whose configurations are the mirror of 
$(2n,1)_0$ and $(2n+1,1)_{-1}$ string webs.
Therefore we see $x_{2n}=x'_{2n}$.

Let us now turn to the case of $m>\frac{\La}{2}$.
It is realized in the IIB background
with a $(1,-1)_{-1}$ 7-brane at $z_1=-\frac{1}{8}\La(\La+8m)$,
a $(1,1)_1$ 7-brane at $z_2=-\frac{1}{8}\La(\La-8m)$ and
two $(1,0)_1$ 7-branes at $z_3=z_4=\frac{1}{8}(\La^2+8m^2)$
following the conventions of \cite{BF3}.
Notice that three 7-branes are mutually non-local since their $(p,q)$ 
charges are distinct in contrast of $m<\La/2$.
The branch cuts are drawn from $z=z_i$ to $z=-\infty$ along the real axis. 
Eq.(\ref{eqn:in}) yields the self-intersection number,
\[
(J\cdot J)=-\frac{1}{2}\left[
        (n_1-n_2-n_3-n_4)^2+(n_1-n_2)^2+(n_3-n_4)^2\right].
\]
Then the possible values of $n_i$ for the BPS webs are only those listed
in Table \ref{tbl:nf=2l}.
\begin{table}
\hspace{2.3cm}
\begin{tabular}{|c|c|c|c|}\hline
        &$b$     &$(n_1,n_2,n_3,n_4)$ or $(n_1,n_2,n_4,n_3)$    
                                         &$(p,q)$\\\hline
(i)     &$2$     &$\pm(1,1,0,0)$         &$\pm(2,0)_0$\\
(ii)    &$1$     &$\pm(n,n-1,0,0)$       &$\pm(2n-1,-1)_{-1}$\\
(iii)   &$1$     &$\pm(n,n-1,1,1)$       &$\pm(2n+1,-1)_1$\\
(iv)    &$1$     &$\pm(n,n-1,1,0)$       &$\pm(2n,-1)_0$\\
(v)     &$1$     &$\pm(n,n,1,0)$         &$\pm(2n+1,0)_1$\\\hline
\end{tabular}
\caption{The possible values of $n_i$ for BPS string webs in $SU(2)$ theory
with massive $N_f=2$ ($m>\frac{\La}{2}$).}
\label{tbl:nf=2l}
\end{table}
When we take the limit $m\rightarrow\infty$ where $z_3=z_4\rightarrow\infty$,
the string webs $n_3,n_4\neq 0$ are decoupled from the world volume theory
because their masses become infinite.
Removing the cuts from $z=z_3$ and $z_4$,
we are left with the spectrum of Yang-Mills theory.

\begin{figure}
\psfrag{Z1}[][]{$z_1$}
\psfrag{Z2}[][]{$z_2$}
\psfrag{Z3}[][]{$z_3$}
\psfrag{Z1st}[][]{$(-1,-1)_{-1}$}
\psfrag{Z2st}[][]{$(1,1)_1$}
\psfrag{Z3st}[][]{$(0,1)_0$}
\psfrag{TT1}[][]{(A) $\phi=\pi/16$}
\psfrag{TT2}[][]{(B) $\phi=0.14\pi$}
\psfrag{TT3}[][]{(C) $\phi=3\pi/16$}
\psfrag{TT4}[][]{$\phi=7\pi/16$}
\includegraphics[width=15.5cm]{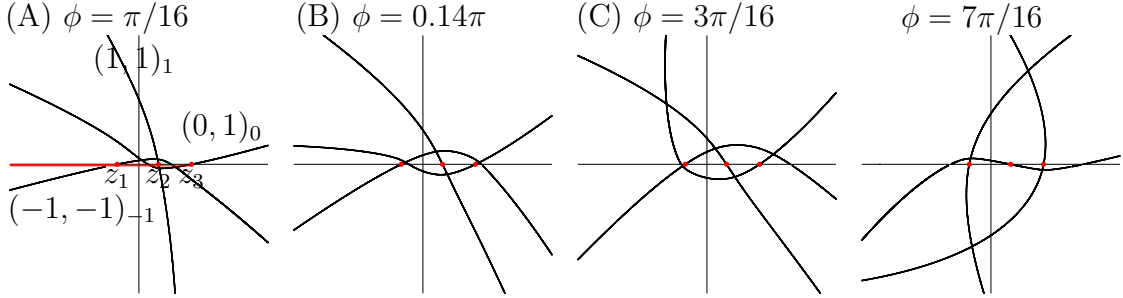}
\caption{BPS strings for $\alpha_i<\phi<\alpha_{i+1}$ ($i=0,1,2$) with
$\alpha_0=0$ and $\alpha_3=\pi/2$.
We set $m/\La=2.5$ for which $\alpha_1=\pi/8$ and $\alpha_2=0.17\pi$.}
\label{fig:HWl}
\end{figure}

The strings from the 7-branes are calculated numerically.
The results are depicted in Fig.\ref{fig:HWl}A-C where the strings in the 
opposite directions have been drawn simultaneously.
The configurations for $\pi/2<\phi<\pi$ are derived by them using 
the relation between $(a(\bar{z}),a_D(\bar{z}))$ and 
$(\bar{a}(\bar{z}),\bar{a}_D(\bar{z}))$.
Now applying the procedure we have used for $m<\La/2$ generate the 
configurations of BPS string webs from them and reproduces the field theory
result.
Since our analysis is almost identical with the case of $m<\La/2$ we 
shall refrain from giving details.
What is different from the $m<\La/2$ case is the observation that
the string webs contain at most three vertices.
In addition, since 7-branes are all mutually non-local the Hanany-Witten 
effect occurs more times than the previous $m<\La/2$ case.
This is in accordance with the field theory result as depicted in Fig.13 of 
\cite{BF3} 
in which there are two types of CMS for $(2n,1)_0$ and 
$(2n-1,1)_{-1}$ string webs with $n\geq 2$ in the upper half-plane.

\section{Conclusions}

We have obtained explicitly the configurations of the 
BPS string webs in the 7-brane background
corresponding to four-dimensional $N=2$ $SU(2)$ theory with $N_f=1$ massless
and $N_f=2$ massive fundamental matters.
The BPS spectrum we have derived is in agreement with
the known results of the field theories.
In the present analysis there have appeared the string webs containing 
four vertices at most.
String webs containing more vertices would appear when we 
consider $SU(2)$ theory with matters of arbitrary masses.

There are other interesting theories realized on the world volume of 
the D3-brane probe, such as
the theory with $E_n$ global symmetry \cite{GZ} \cite{MN}-\cite{I} 
and the theory with $Sp(2n)$ gauge symmetry \cite{DLS}-\cite{AFM},
whose BPS spectrum is unknown.
It is worth analyzing the BPS string webs in corresponding backgrounds.

\section*{Acknowledgements}

I would like to thank S.-K.Yang for useful discussions and the careful
reading of the manuscript.

\section*{Appendix A: Explicit forms of $a(z)$ and $a_D(z)$}

\renewcommand{\theequation}{A.\arabic{equation}}
\setcounter{equation}{0}

The Seiberg-Witten curve for $SU(2)$ theory with $N_f=1$ massless matter is
\[
y^2=x^3-zx^2-\frac{\La^6}{64}.
\]
The Seiberg-Witten periods $a(z)$ and $a_D(z)$ in the conventions we use in the
text are given by \cite{BF2},
\beqa
a(z)&=&\frac{1}{2}\sqrt{2z}
        F\left(-\frac{1}{6},\frac{1}{6},1;-\frac{1}{z^3}\right),\CR
a_D(z)&=&\left\{\begin{array}{cc}
e^{-i\pi/3}f_D(z)-a(z)  &\mbox{for}\; 0\leq \mbox{arg}z<2\pi/3,\\
f_D(z)-2a(z)            &\mbox{for}\; 2\pi/3\leq \mbox{arg}z<\pi,\\
-f_D(z)+a(z)            &\mbox{for}\; \pi\leq \mbox{arg}z<4\pi/3,\\
e^{-2i\pi/3}f_D(z)      &\mbox{for}\; 4\pi/3\leq \mbox{arg}z<2\pi,
\end{array}\right.
\label{eqn:aad1}
\eeqa
where we set $\La^6=256/27$, $F$ is the standard hypergeometric function and 
\[
f_D(z)=\frac{\sqrt{2}}{12}(z^3+1)F\left(\frac{5}{6},\frac{5}{6},2;1+z^3\right).
\]

The Seiberg-Witten curve of $SU(2)$ theory with $N_f=2$ massive matters is
\[
y^2=x^2(x-z)-\frac{\La^4}{64}(x-z)+\frac{\La^2}{4}m_1m_2x-\frac{\La^4}{64}
(m_1^2+m_2^2),
\]
which may be rewritten in the {Weirstra\ss} 
form $y^2=x^3+f(z)x+g(z)$.
Setting $m_1=m_2\equiv m$, we have \cite{BF3}
\beqa
a(z)&=&f^{(1)}(z)+\frac{m}{\sqrt{2}}, \CR
a_D(z)&=&\left\{\begin{array}{cc}
        f^{(2)}(z)      &\mbox{for}\; \mbox{Im}z\geq 0,\\
        -f^{(2)}(z)     &\mbox{for}\; \mbox{Im}z<0,
        \end{array}\right. \label{eqn:aad2}
\eeqa
where $f^{(i)}(z)$ ($i=1,2$) are given by 
\[
f^{(i)}(z)
=\frac{\sqrt{2}}{4\pi}\left[\frac{4}{3}zI_1^{(i)}(z)-2I_2^{(i)}(z)
        -\frac{\La^2}{2}m^2I_3^{(i)}(z)\right].
\]
Here $I_j^{(1)}(z)$ ($i=1,2,3$) are obtained as
\beqa
I_1^{(1)}(z)&=&\frac{2}{(e_1-e_3)^{1/2}}K(k),\CR
I_2^{(1)}(z)&=&\frac{2}{(e_1-e_3)^{1/2}}\left[e_1K(k)+(e_3-e_1)E(k)\right],\CR
I_3^{(1)}(z)&=&\frac{2}{(e_1-e_3)^{3/2}}
        \left[\frac{1}{1-\tilde{c}+k'}K(k)
        +\frac{4k'}{1+k'}\frac{1}{(1-\tilde{c})^2+k'^2}
        \Pi\left(\nu (\tilde{c}), \frac{1-k'}{1+k'}\right)\right],\nonumber
\eeqa
where
\beqa
e_1&=&\frac{z}{6}-\frac{\La^2}{16}
+\frac{1}{2}\sqrt{\left(z+\frac{\La^2}{8}+\La m\right)
        \left(z+\frac{\La^2}{8}-\La m\right)},\CR
e_2&=&-\frac{z}{3}+\frac{\La^2}{8},\CR
e_3&=&\frac{z}{6}-\frac{\La^2}{16}
	-\frac{1}{2}\sqrt{\left(z+\frac{\La^2}{8}+\La m\right)
        \left(z+\frac{\La^2}{8}-\La m\right)},\CR
k^2&=&\frac{e_2-e_3}{e_1-e_3},\CR
k'^2&=&1-k^2=\frac{e_2-e_1}{e_3-e_1},\CR
\tilde{c}&=&\frac{-\frac{z}{3}-\frac{\La^2}{8}-e_3}{e_1-e_3},\CR
\nu(\tilde{c})&=&\left(\frac{1-\tilde{c}+k'}{1-\tilde{c}-k'}\right)^2
                \left(\frac{1-k'}{1+k'}\right)^2,\nonumber
\eeqa
and $I_j^{(2)}(z)$ are obtained from $I_j^{(1)}(z)$ by exchanging $e_1$ 
and $e_3$.
$K(k)$, $E(k)$ and $\Pi(\nu,k)$ are the three standard elliptic integrals 
defined by
\beqa
K(k)&=& \int_{0}^{1}\frac{dx}{\sqrt{(1-x^2)(1-k^2x^2)}},\CR
E(k)&=& \int_{0}^{1}dx\sqrt{\frac{1-k^2x^2}{1-x^2}},\CR
\Pi(\nu,k)&=& \int_{0}^{1}\frac{dx}{\sqrt{(1-x^2)(1-k^2x^2)}(1-\nu x^2)}.
        \nonumber
\eeqa

\newpage


\begin{thebibliography}{99}

\bibitem{SeWi} N. Seiberg and E. Witten,
Nucl. Phys. {\bf B426} (1994) 19, hep-th/9407087;
 Nucl. Phys. {\bf B431} (1994) 484,
 hep-th/9408099.

\bibitem{BF2} A. Bilal and F. Ferrari,
Nucl. Phys. {\bf B469} (1996) 387,
 hep-th/9602082;
Nucl. Phys. {\bf B480} (1996) 589,
 hep-th/9605101.

\bibitem{BF3} A. Bilal and F. Ferrari,
Nucl. Phys. {\bf B516} (1998) 175,
 hep-th/9706145.

\bibitem{Sen} A. Sen, 
Nucl. Phys. {\bf B475} (1996) 562,
 hep-th/9605150.

\bibitem{BDS} T. Banks, M.Douglas and N. Seiberg,
Phys. Lett. {\bf B387} (1996) 278,
 hep-th/9605199.

\bibitem{Fy} A. Fayyazuddin,
Nucl. Phys. {\bf B497} (1997) 101,
 hep-th/9701185.

\bibitem{BFy} O. Bergman and A. Fayyazuddin,
Nucl. Phys. {\bf B531} (1998) 108,
 hep-th/9802033.

\bibitem{MNS} A. Mikhailov, N. Nekrasov and S.Sethi,
Nucl. Phys. {\bf B531} (1998) 345,
 hep-th/9803142.

\bibitem{GZ} M. R. Gaberdiel and B. Zwiebach,
Nucl. Phys. {\bf B518} (1998) 151,
 hep-th/9709013.

\bibitem{KL} M. Krogh and S. Lee,
Nucl. Phys. {\bf B516} (1998) 241, 
 hep-th/9712050.

\bibitem{MO} Y. Matsuo and K. Okuyama,
Phys. Lett. {\bf B426} (1998) 294,
 hep-th/9712070.

\bibitem{KS} I. Kishimoto and N. Sasakura,
Phys. Lett. {\bf B432} (1998) 305, 
 hep-th/9712180.

\bibitem{dWHIZ} O. DeWolfe, T. Hauer, A. Iqbal and B. Zwiebach,
Nucl. Phys. {\bf B534} (1998) 261,
 hep-th/9805220.

\bibitem{dWZ} O. DeWolfe and B. Zwiebach, 
 {\it String Junctions for Arbitary Lie Algebra Representations},
 hep-th/9804210.

\bibitem{Sen3} A. Sen, 
Phys. Rev. {\bf D55} (1997) 2501,
 hep-th/9608005.

\bibitem{Sen2} A. Sen, 
 {\it String Network},
 hep-th/9711130.

\bibitem{MN} J.A. Minahan and D. Nemeschansky,
Nucl. Phys. {\bf B482} (1996) 142,
 hep-th/9608047; 
Nucl. Phys. {\bf B489} (1997) 24,
 hep-th/9610076.

\bibitem{S} N. Seiberg,
Phys. Lett. {\bf B388} (1996) 753,
 hep-th/9608111;
Phys. Lett. {\bf B390} (1997) 169,
 hep-th/9609161.

\bibitem{I} Y. Imamura,
Phys. Rev. {\bf D58} (1998) 106005,
 hep-th/9802189.

\bibitem{DLS} M.R. Douglas, D.A. Lowe and J.H. Schwarz,
Phys. Lett. {\bf B394} (1997) 297,
 hep-th/9612062.

\bibitem{Iq} A. Iqbal,
{\it Self-Intersection Number of BPS Junctions in Backgrounds of Three and Seven-Branes},
 hep-th/9807117.

\bibitem{FS} A. Fayyazuddin and M. Spalinski,
Nucl. Phys. {\bf B535} (1998) 219,
 hep-th/9805096.

\bibitem{AFM} O. Aharony, A. Fayyazuddin and J. Maldacena,
J. High Energy Phys. {\bf 9807} (1998) 013,
 hep-th/9806159.

\end{thebibliography}
\end{document}